18 March 2026

# Eliminating Infinite Self-Energies From Classical Electrodynamics

by Andrew T. Hyman[1]

## ABSTRACT

The theory of point-particles in classical electrodynamics has a well-known problem of infinite self-energy, and the same is true of quantum electrodynamics. Instead of concluding that there is no such thing as a true point-particle, it is shown here how to remove the infinities by supposing that the electromagnetic field tensor has a symmetric part. This does not change the physics in flat spacetime, as the equation of motion and the antisymmetric part of the retarded fields appearing in the equation of motion are unaffected. The symmetric part of the field tensor is not observable and therefore it need not be gauge-invariant, whereas the antisymmetric part is observable, gauge-invariant, and satisfies both the Maxwell Equations and the field equations governing the whole electromagnetic field tensor. This approach goes well beyond prior efforts at classical renormalization, and also entails a new derivation of the Lorentz-Abraham-Dirac (LAD) equation of motion. Implications for General Relativity are discussed in the Appendix.

## I. Introduction

The usual renormalization procedure in quantum electrodynamics to remove infinities is somewhat artificial, and essentially the same problem exists in classical electrodynamics.[2] This predicament has suggested to some physicists that no true point-charges can exist in either the classical or the quantum theory.[3] But, as David Griffiths has put it, "in neither theory do we know how to avoid the point particle as a theoretical construct."[4] So, the present article discusses the classical problem and presents a new way to deal with the infinities, without altering any physical predictions, and without avoiding the point particle as a theoretical construct.

---

[1] The Institute for Intermediate Study (www.tifis.org). EMAIL: andrewthyman@gmail.com.

[2] Alhassan M and Alharbi F 2026 Radiation reaction: status and recent developments *Eur. Phys. J. Plus* **141** 211.

[3] Steane A 2015 Tracking the radiation reaction energy when charged bodies accelerate *Am. J. Phys* **83** 703.

[4] Griffiths D 2020 *Introduction to Elementary Particles* (2nd Revised Edition, Wiley & Sons) 219, 220.

As shown here, it turns out that the infinite self-energy problem can be solved by employing an electromagnetic field tensor that has both an antisymmetric part and a symmetric part. That innovation eliminates the infinities, but does not affect the motion of point-particles in the context of Special Relativity. In the context of relativistic gravitation, if the new formula for the electromagnetic energy-momentum tensor is a source-term for gravitation,[5,6] then the symmetric part of the electromagnetic field may indeed become observable (i.e. less ghostly).

The usual field is antisymmetric:

$$F^{\alpha\lambda} + F^{\lambda\alpha} = 0 \;. \tag{1.1}$$

This field satisfies the Maxwell Equations in free space (i.e. where no point-charges are located):

$$F_{\alpha\beta,\lambda} + F_{\lambda\alpha,\beta} + F_{\beta\lambda,\alpha} = 0 \;. \tag{1.2}$$

$$F^{\alpha\lambda}{}_{,\lambda} = 0 \;. \tag{1.3}$$

Repeated indices are summed from zero to three, commas denote partial differentiation with respect to the four spacetime coordinates $x^{\mu}$, and indices are raised and lowered by contraction with the Minkowski tensor (eta) whose non-zero components are all diagonal (-1,1,1,1). The usual formula for electromagnetic energy-momentum in free space is this:

$$T^{\alpha\beta} = \frac{1}{4\pi}\left[ F^{\alpha}{}_{\lambda} F^{\beta\lambda} - \frac{1}{4}\eta^{\alpha\beta} F^{\lambda\sigma} F_{\lambda\sigma} \right]. \tag{1.4}$$

Equations (1.2) and (1.3) ensure that the four-divergence of (1.4) vanishes in free space.

[5] Burinskii A 2008 The Dirac-Kerr-Newman electron *Grav. & Cosmology,* **14** 109.
[6] Holzhey C and Wilczek F 1992 Black holes as elementary particles *Nuclear Physics B,* **380** 447.

As described above, the usual formulation of classical electrodynamics in free space involves six field components of an antisymmetric tensor "F" (i.e. $F_{01}$, $F_{02}$, $F_{03}$, $F_{12}$, $F_{23}$, $F_{31}$) governed by eight field equation components. Instead consider an alternative formulation that involves 15 field components of a tensor "Φ" that is not antisymmetric:

$$\Phi^{\lambda}{}_{\lambda}=0\,. \tag{1.6}$$

$$\Phi^{\alpha\lambda}{}_{,\lambda}=0\,. \tag{1.7}$$

$$\Phi_{\alpha\beta,\lambda}-\Phi_{\alpha\lambda,\beta}=0\,. \tag{1.8}$$

The field equations (1.6) through (1.8) in free space are similar to the usual equations (1.1) through (1.3) in that they are linear, and there are no partial derivatives higher than the first order.

By supposing that the electromagnetic field has both symmetric and antisymmetric components, the number of field components (i.e. 16) is not changed, but their formulae are changed, and the diagonal components no longer vanish. Occam's razor does counsel against constructing explanations using more than the smallest number of elements, but infinite self-energies have heretofor been a huge and problematic element that is eliminated here. We can construct a conserved symmetric energy-momentum tensor that is similar to (1.4) in which there are no partial derivatives:

$$T^{\alpha\beta}=\frac{1}{4\pi}\left[\Phi^{\lambda\alpha}\Phi_{\lambda}{}^{\beta}-\frac{1}{2}\eta^{\alpha\beta}\Phi^{\sigma\lambda}\Phi_{\sigma\lambda}\right]+\frac{n}{2\pi}\left[2(\Phi^{\alpha\lambda}\Phi_{\lambda}{}^{\beta}+\Phi^{\beta\lambda}\Phi_{\lambda}{}^{\alpha})-\eta^{\alpha\beta}\Phi^{\sigma\lambda}\Phi_{\lambda\sigma}-2\Phi^{\alpha\lambda}\Phi^{\beta}{}_{\lambda}\right]. \tag{1.9}$$

Here "n" is a dimensionless constant, and throughout this article "n" (with or without a subscript) will represent a dimensionless constant. Equations (1.6) through (1.8) ensure that the four-divergence of (1.9) vanishes in free space.

If the dimensionless constant "n" in equation (1.9) is set to -1/4 then it is straightforward to prove that the new energy-momentum formula (1.9) reduces to the old formula (1.4). One can prove this by decomposing the phi-field into its antisymmetric and symmetric parts, and eliminating "F" from (1.4) using the antisymmetric formula:

$$F_{\alpha\beta} = \Phi_{\alpha\beta} - \Phi_{\beta\alpha} \,. \tag{1.10a}$$

$$W_{\alpha\beta} = \Phi_{\alpha\beta} + \Phi_{\beta\alpha} \,. \tag{1.10b}$$

Thus the sum of the "F" and "W" tensors is twice the phi tensor. To eliminate infinite self-energy, as will be shown in the next section, the proper value of "n" is unique:

$$n=-1/8. \tag{1.11}$$

The usual well-known retarded fields of a point-charge are as follows:

$$F_{\lambda\beta} = qR_\beta\left[\frac{\ddot{z}_\lambda}{\varepsilon^2} + \frac{\dot{z}_\lambda}{\varepsilon^3}(R^\mu \ddot{z}_\mu + 1)\right] - qR_\lambda\left[\frac{\ddot{z}_\beta}{\varepsilon^2} + \frac{\dot{z}_\beta}{\varepsilon^3}(R^\mu \ddot{z}_\mu + 1)\right]. \tag{1.12}$$

Overdots denote differentiation with respect to proper time, "z" represents a particle's position, and coordinates are chosen so that the speed of light is unity (c=1). The retarded formula for the new fields is this:

$$\Phi_{\lambda\beta} = \frac{-q\dot{z}_\lambda \dot{z}_\beta}{\varepsilon^2} + qR_\beta\left[\frac{\ddot{z}_\lambda}{\varepsilon^2} + \frac{\dot{z}_\lambda}{\varepsilon^3}(R^\mu \ddot{z}_\mu + 1)\right]. \tag{1.13}$$

The retardation condition is as usual:

$$R_\lambda R^\lambda = 0. \tag{1.14}$$

$$R_\lambda \equiv x_\lambda - z_\lambda(s). \tag{1.15}$$

$$\varepsilon \equiv -R_\lambda \dot{z}^\lambda. \tag{1.16}$$

Equation (1.12) can be obtained by simply putting (1.13) into (1.10).

It is shown below that this reformulation of classical electrodynamics allows point-particles to exist quite comfortably. This without changing their motion at all, because the antisymmetric part of the retarded fields, and the equation of motion which is re-derived here, all remain the same.

## II. Introducing a four-potential and taking the static limit

The field equation (1.8) indicates that we can introduce a four-potential:

$$\Phi_{\lambda\beta} = A_{\lambda,\beta}. \qquad (2.1)$$

where $A_\lambda$ is the Liénard-Wiechert (L-W) potential:

$$A_\lambda = \frac{q\dot{z}_\lambda}{R^\sigma \dot{z}_\sigma}. \qquad (2.2)$$

In the static limit, only the zeroth component of (2.2) is nonvanishing. Let us now consider the zero-zero component of the self-energy-momentum; that component is the self-energy density in free space. Putting (2.1) into (1.9), and taking the static limit gives this formula for self-energy density:

$$T^{00} = \frac{1}{8\pi}[1+8n]\left[A^{0,m} A_{0,m}\right]. \qquad (2.3)$$

Equation (1.11) thus ensures that the self-energy-density in free space caused by a static particle vanishes, and thus the total self-energy in free space caused by any single point-charge must be finite. In other words, the requirement of finite self-energy has led us uniquely to an appropriate formula for electromagnetic energy-momentum more generally. We have not yet considered the mechanical energy that is located at the point-particle itself. An appropriate formula for mechanical energy-momentum is obtained in the next section, to complement the new free-space energy-momentum formula (1.9).

## III. Flux of energy-momentum leads to mechanical energy-momentum formula

The technique for calculating the flux of energy-momentum from a point-charge is well-known, by surrounding a point-source with a tiny surface of integration called a Bhabha tube.[7] If we calculate the flux of the energy-momentum (1.9) from a point-charge, given (1.11), then this is the result:

$$f^{\lambda}=q[\Phi^{\omega\lambda}-\Phi^{\lambda\omega}]\dot{z}_{\omega}+\frac{q}{2}\frac{dA^{\lambda}}{ds}+\frac{2q^{2}}{3}[\dot{z}^{\lambda}\ddot{z}^{\mu}\ddot{z}_{\mu}]. \quad (3.1)$$

On the right-hand-side, the fields and the potentials are both external only, meaning that they are sums of the individual fields and potentials caused by all other particles, not including the fields and potentials caused by the particle from which the flux has been calculated. Conservation of energy-momentum demands that this flux (3.1) be balanced by the change in mechanical energy-momentum ($p^{\lambda}$) of the particle from which the flux has been calculated:

$$f^{\lambda}+\dot{p}^{\lambda}=0. \quad (3.2)$$

Equations (3.1) and (3.2) suggest that the mechanical energy-momentum is given this way:

$$p^{\lambda}=m\dot{z}^{\lambda}-\frac{q}{2}A^{\lambda}-\frac{2q^{2}}{3}\ddot{z}^{\lambda}. \quad (3.3)$$

The last term on the right-hand-side of (3.3) is necessary so that equations (3.1) and (3.2) ensure we have a sensible equation of motion that contracts with four-velocity to yield zero:

$$0=q[\Phi^{\omega\lambda}-\Phi^{\lambda\omega}]\dot{z}_{\omega}+\frac{2q^{2}}{3}[\dot{z}^{\lambda}\ddot{z}^{\mu}\ddot{z}_{\mu}]+m\ddot{z}^{\lambda}-\frac{2q^{2}}{3}\dddot{z}^{\lambda}. \quad (3.4)$$

Equation (3.4) is none other than the famous Lorentz-Abraham-Dirac (LAD) equation of motion, as is clear from equation (1.10). The Lorentz covariant equation (3.4) was, despite its name, apparently first written by Wolfgang Pauli in 1921.[8] The present incidental derivation of this equation is believed to be entirely new.

[7] Bhabha HJ and Corben HC 1941 General classical theory of spinning particles in a Maxwell field *Proc. Roy. Soc. London,* **178** 273.

Equations (1.13) and (3.4) completely describe the motion and interaction of a set of classical point charges, assuming they have no dipole moments. This motion and interaction are the same using the old energy-momentum (1.4) as the new energy-momentum given by (1.9) with (1.11).

## IV. Energy of a set of static particles

We have already seen in Section II that the energy of a set of static particles vanishes in free space, according to the new energy-momentum given by (1.9) and (1.11). Thus, the total energy of a set of N static particles must be simply the sum of their mechanical energies which can be obtained from (3.3):

$$\text{Energy}=\sum_{i=1}^{N}\left[m_i-\frac{q_i}{2}\bullet(A^0)_{\text{at particle i}}\right]. \tag{4.1}$$

Next, we use the formula (2.2) to rewrite (4.1):

$$\text{Energy}=\left[\sum_{i=1}^{N}m_i\right]+\left[\sum_{i\neq j}\frac{q_iq_j}{r_{ij}}\right]. \tag{4.2}$$

Happily, the formula (4.2) is completely finite. The first bracketed term simply embodies the principle that mass equals energy per Einstein's famous formula (recall we have chosen coordinates so that c=1). The second bracketed term was well-understood decades before the Maxwell Equations were devised.[9] The new energy-momentum formula (1.9) with (1.11) has led to the usual electrostatic formula (4.2), without any infinities.

## V. Radiated energy-momentum and its trace

Consider the last part of equation (1.9):

$$\tau^{\alpha\beta}\equiv\frac{n}{2\pi}\left[2(\Phi^{\alpha\lambda}\Phi_{\lambda}{}^{\beta}+\Phi^{\beta\lambda}\Phi_{\lambda}{}^{\alpha})-\eta^{\alpha\beta}\Phi^{\sigma\lambda}\Phi_{\lambda\sigma}-2\Phi^{\alpha\lambda}\Phi^{\beta}{}_{\lambda}\right]. \tag{5.1}$$

---

[8] Pauli W 1921 Relativitatstheorie, *Enzyl. Math. Wiss*. **5**, 543; translated in 1958 by G. Field as *Theory of Relativity* (Dover) 99.

[9] Poisson SD 1842 *A Treatise of Mechanics* (Longman) **2** 393.

We are concerned now with radiation, so we are only interested in the part of the energy-momentum that falls off as $1/r^2$, and thus we must find the part of the phi-field that drops off as 1/r. Equation (1.13) tells us this:

$$\Phi^{\lambda\beta}_{\text{rad.}} = qR^{\beta}\left[\frac{\ddot{z}^{\lambda}}{\varepsilon^{2}} + \frac{\dot{z}^{\lambda}}{\varepsilon^{3}}R^{\mu}\ddot{z}_{\mu}\right]. \tag{5.2}$$

Next, we put (5.2) into (5.1), in view of (1.14) and (1.16):

$$\tau^{\alpha\beta} = 0\,. \tag{5.3}$$

This tells us that an accelerating point-charge radiates energy-momentum (1.9) that is independent of the dimensionless constant "n." That is, the radiation is the same regardless of whether we employ the traditional value n=-1/4 or instead the new and improved value given by equation (1.11), n=-1/8. Accordingly, the well-known Larmor formula happens to be preserved by the new energy-momentum formula (1.9) with (1.11).

As for the trace of the electromagnetic energy-momentum tensor, it is generally non-vanishing per (1.9) with (1.11), though that trace does completely vanish in the case of the radiation field (5.2) from a single point-charge. Empirically, conformal invariance is broken, which implies that energy-momentum tensors need not vanish. Regardless, according to Fritz Rohrlich, "the physical situation singled out by this [conformal] invariance property does not seem to be of sufficient simplicity or importance to warrant special attention in a theory with nonvanishing rest masses."[10] Therefore, tracelessness of the energy-momentum tensor is not demanded here.

## VI. Gauge invariance

As can be seen from equation (2.1), the phi-field remains the same when an arbitrary constant vector is added to the four-potential. However, the phi-field does not remain the same when an arbitrary four-gradient is added to the four-potential. So, one could conclude that the present formulation of classical electrodynamics is not gauge invariant. However, the observable part of the phi-field is the antisymmetric part, because that is the only part that enters into the equation of motion (3.4), whereas the symmetric part is not observable and therefore need not be gauge-invariant.[11]

[10] Rohrlich F 2007 *Classical Charged Particles* (World Scientific) 104.

Clearly, gauge invariance has played a central role in modern physics.[12] Oddly enough, it has been known since 1997 that the L-W retarded four-potential (2.2) can be obtained as the divergence of a gauge-invariant "generator" tensor.[13] Therefore, it is not entirely clear how much the principle of gauge invariance demands of classical electrodynamics, or why it does so. In any event, classical electrodynamics is not a quantum theory, and so the requirement of gauge invariance does not seem essential in classical electrodynamics, for the same reason that it is not essential in the Newtonian theory of gravitation. Yet, the new field equations presented here do seem to be gauge invariant, because the observable, antisymmetric part of the field is gauge-invariant.

Various other proposed modifications of classical electrodynamics have discarded the principle of gauge invariance, for example a theory discussed by Paul Davies entailed an energy-momentum tensor that must explicitly rely upon a four-potential.[14] Davies wrote, "electrodynamics is no longer gauge invariant, and the potentials…have physical status." Here there is no such physical status, because the physical results are the same as for the usual formulation of classical electrodynamics with the infinite self-energies. One might say that the present new formulation of classical electrodynamics is a result of performing a sort of classical renormalization on the usual formulation, but differently and more fundamentally than the "classical renormalization" achieved by Dirac in 1938, and by others since then.[15,16]

In the case of energy-momentum, its value in free space has been given by (1.9) together with (1.11), and its mechanical value at the point-charge itself has been given by (3.3). Corresponding formulae for angular momentum are respectively these:

---

[11] Levin FS 2002 *An Introduction to Quantum Theory* (Cambridge U. Press) 469.
[12] Jackson JD & Okun LB 2001 Historical roots of gauge invariance. *Rev. Mod. Phys.* **73** 663-680.
[13] López-Bonilla JL, Ovando G & Rivera JM 1997 Generators for 4-potential and the Faraday's tensor of the Liénard-Wiechert field. *Indian J. Pure Appl. Math*. **28** 1355-1360.
[14] Davies PCW 1971 Charged-particle creation in cosmology, *Nuovo Cim*. **6B** 164-178.
[15] Dirac PAM 1938 Classical Theory of Radiating Electrons *Proc. Roy. Soc. A.* **167** 148.
[16] Norton A 2009 The altenative to classical mass renormalization for tube-based self-force calculations *Classical and Quantum Grav.* **26** 105009.

$$\Theta^{\alpha\beta\lambda}=x^{\alpha}T^{\beta\lambda}-x^{\beta}T^{\alpha\lambda}-\frac{1}{8\pi}\Big[A^{\alpha}A^{\lambda,\beta}-A^{\beta}A^{\lambda,\alpha}+A^{\lambda}A^{\alpha,\beta}-A^{\lambda}A^{\beta,\alpha}-A_{\sigma}A^{\alpha,\sigma}\eta^{\beta\lambda}+A_{\sigma}A^{\beta,\sigma}\eta^{\alpha\lambda}$$

$$+(A^{\alpha}A^{\beta,\lambda}-A^{\beta}A^{\alpha,\lambda})+A_{\sigma}(A^{\sigma,\alpha}\eta^{\beta\lambda}-A^{\sigma,\beta}\eta^{\alpha\lambda})\Big]. \qquad (6.1)$$

$$\omega^{\alpha\beta}=z^{\alpha}p^{\beta}-z^{\beta}p^{\alpha}. \qquad (6.2)$$

Putting aside the issue of gauge invariance (which is not a problem so long as the equation of motion does not depend upon any fields other then the usual antisymmetric one), these last two equations satisfy all of the usual requirements.[17]

## VII. Conclusion

The mathematics of classical point-charges can be modified to avoid infinite self-energies, without modifying any experimental predictions in the context of Special Relativity; the retarded fields and the equation of motion all remain the same. The new field equations are consistent with General Relativity or any other metric theory of gravitation, as discussed in the Appendix.

The new field equations and the energy-momentum formula presented here do not require use of any potentials, just like the old formulae. The observable part of the electromagnetic field strength tensor remains gauge invariant and satisfies the usual Maxwell Equations in free space, whereas the symmetric part of that tensor is not gauge invariant; this is acceptable because the symmetric part is not observable, in the sense that it does not appear in the equation of motion. The Lorentz-Abraham-Dirac equation of motion for a point-charge has also been derived here, in conjunction with finding a more appropriate formula for energy-momentum that does not implausibly require infinities.

The reformulation of classical electrodynamics described in this article does not address the well-known problem of runaway solutions due to radiation reaction. Consequently, even if there is merit in the present reformulation, David Jackson is still correct: “a completely satisfactory classical treatment of the reactive effects of radiation does not exist.”[18]

---

[17] Teitelboim C, Villarroel D, and van Weert C 1980 Classical Fields of Retarded Fields and Point Particles. *Riv. Nuovo Cim*. **3**, 1.

[18] Jackson JD 1998 *Classical Electrodynamics*, 3d ed. (Wiley & Sons) 745.

There is apparently no compelling physical reason to prefer that the dimensionless constant “n” in equation (1.9) should be given by the usual value (-1/4) instead of by the proposed value (-1/8). The advantage in the latter case is that there are no infinite self-energies according to classical electrodynamics. The latter case generally does give a non-vanishing trace of the energy-momentum tensor, but for both values of “n” the motion and interaction of point-particles have been shown to be the same in Special Relativity, in both cases the trace vanishes if we only consider radiation from a point-source, and in neither case is the principle of conformal invariance strong enough to require tracelessness.

Infinite self-energy of a point-charge is similar to the apparent singularity at the Schwarzschild radius in General Relativity. That Schwarzschild singularity is merely an artifact of the chosen coordinate system, rather than a genuine physical singularity as had long been mistakenly believed.[19] Likewise, the infinite self-energy of a point charge seems to be merely an artifact of assigning particular values to the dimensionless constant in equation (1.9), and changing that constant’s entrenched value — as has been done here — does not change the motion or interaction of classical point-charges. The unobservable symmetric part of the electromagnetic field tensor, introduced here for Special Relativity, is analogous to a coordinate condition in General Relativity; neither is observable, and neither is gauge invariant.

## VIII. Acknowledgments

I would like to acknowledge helpful comments along the winding road to this article, from David Griffiths, Jeffrey Bub, Eric Poisson, Kirk McDonald, and José Luis López-Bonilla. This is not to say that they all agree or disagree with what is written here, and of course any errors are mine alone.

## Appendix: Extension to General Relativity

The proposed field equations (1.6) through (1.8) might be generalized to fit within any metric theory of gravitation, such as General Relativity, by simply switching commas with semicolons. That switch signifies covariant derivatives instead of simple partial derivatives:

$$\Phi^{\lambda}{}_{\lambda} = 0 \quad . \qquad \text{(A.1)}$$

$$\Phi^{\alpha\lambda}{}_{;\lambda} = 0 \quad . \qquad \text{(A.2)}$$

[19] Penrose R 2017 *Fashion, Faith, and Fantasy in the New Physics of the Universe* (Princeton U. Press) 233.

$$\Phi_{\alpha\beta;\lambda} - \Phi_{\alpha\lambda;\beta} = 0 \;. \qquad \text{(A.3)}$$

It turns out that these electromagnetic field equations would imply a substantial constraint upon spacetime curvature. To see that this is so, let us take the first covariant derivatives of (A.3):

$$\Phi_{\alpha\beta;\lambda;\mu} - \Phi_{\alpha\lambda;\beta;\mu} = 0 \;. \qquad \text{(A.4)}$$

Therefore:

$$[\Phi_{\alpha\beta;\lambda;\mu} - \Phi_{\alpha\lambda;\beta;\mu}] - [\Phi_{\alpha\mu;\lambda;\beta} - \Phi_{\alpha\lambda;\mu;\beta}] - [\Phi_{\alpha\beta;\mu;\lambda} - \Phi_{\alpha\mu;\beta;\lambda}] = 0 \;. \qquad \text{(A.5)}$$

Rearranging terms in (A.5) gives this:

$$[\Phi_{\alpha\beta;\lambda;\mu} - \Phi_{\alpha\beta;\mu;\lambda}] + [\Phi_{\alpha\lambda;\mu;\beta} - \Phi_{\alpha\lambda;\beta;\mu}] + [\Phi_{\alpha\mu;\beta;\lambda} - \Phi_{\alpha\mu;\lambda;\beta}] = 0 \;. \qquad \text{(A.6)}$$

Each bracketed term in (A.6) would vanish if the equality of mixed partial derivatives were applicable to covariant derivatives, but it is not applicable. Instead, we would get this curvature constraint:[20]

$$\Phi^{\sigma}{}_{\beta} R_{\alpha\sigma\lambda\mu} + \Phi^{\sigma}{}_{\lambda} R_{\alpha\sigma\mu\beta} + \Phi^{\sigma}{}_{\mu} R_{\alpha\sigma\beta\lambda} = 0 \;. \qquad \text{(A.7)}$$

Equation (A.7) would be an extensive constraint upon curvature. However, the Equivalence Principle does not necessarily apply to conditions that are not observable in flat spacetime, so the properties of unobservable quantities in Special Relativity may not all need to hold true in locally inertial coordinates. Of course, the LAD equation of motion involves only the antisymmetric part of the fields, which is gauge-invariant unlike the symmetric unobservable part. Beyond these remarks on curved spacetime, it seems premature to venture.

---

[20] Weinberg S 1972 *Gravitation and Cosmology* (Wiley & Sons) 140-141.